\documentclass[prc,twocolumn,secnumroman,tightenlines
,amssymb,aps,nobibnotes,superscriptaddress,showpacs,balancelastpage,floatfix]{revtex4}
\usepackage{docs}
\usepackage{graphicx}
\usepackage{multirow}           
\usepackage{color}	
\expandafter\ifx\csname package@font\endcsname\relax\else
\expandafter\expandafter
\expandafter\usepackage
\expandafter{\csname package@font\endcsname}
\fi
\DeclareRobustCommand\substyle{\name@idx{document substyle}}
\DeclareRobustCommand\classoption{\name@idx{document class option}}
\DeclareRobustCommand\classname{\name@idx{document class}}
\def\name@idx#1#2{{\ttfamily#2}
\index{#2\space#1=\string\ttt{#2}\space#1}\index{#1>#2=\string\ttt{#2}}}

\begin{document}
\title{Decay of excited nuclei produced in $^{78,82}$Kr + $^{40}$Ca reactions at 5.5 MeV/nucleon}
\author{G.~Ademard}
\affiliation {Grand Acc\'el\'erateur National d'Ions Lourds (GANIL), CEA/DSM-CNRS/IN2P3, Boulevard H. Becquerel, F-14076, Caen, France}
\author{J.P.~Wieleczko}
\email{wieleczko@ganil.fr}
\affiliation {Grand Acc\'el\'erateur National d'Ions Lourds (GANIL), CEA/DSM-CNRS/IN2P3, Boulevard H. Becquerel, F-14076, Caen, France}
\author{J.~Gomez del Campo}
\affiliation{Physics Division, Oak Ridge National Laboratory, Oak Ridge, TN 37831, USA}
\author{M.~La Commara}
\affiliation{Dipartimento di Scienze Fisiche, Universit\`a di Napoli "Federico II", I-80126, Napoli, Italy}
\affiliation{INFN, Sezione di Napoli, I-80126, Napoli, Italy}
\author{E.~Bonnet}
\affiliation {Grand Acc\'el\'erateur National d'Ions Lourds (GANIL), CEA/DSM-CNRS/IN2P3, Boulevard H. Becquerel, F-14076, Caen, France}
\author{M.~Vigilante}
\affiliation{Dipartimento di Scienze Fisiche, Universit\`a di Napoli "Federico II", I-80126, Napoli, Italy}
\affiliation{INFN, Sezione di Napoli, I-80126, Napoli, Italy}
\author{A.~Chbihi}
\affiliation {Grand Acc\'el\'erateur National d'Ions Lourds (GANIL), CEA/DSM-CNRS/IN2P3, Boulevard H. Becquerel, F-14076, Caen, France}
\author{J.D.~Frankland}
\affiliation {Grand Acc\'el\'erateur National d'Ions Lourds (GANIL), CEA/DSM-CNRS/IN2P3, Boulevard H. Becquerel, F-14076, Caen, France}
\author{E.~Rosato}
\affiliation{Dipartimento di Scienze Fisiche, Universit\`a di Napoli "Federico II", I-80126, Napoli, Italy}
\affiliation{INFN, Sezione di Napoli, I-80126, Napoli, Italy}
\author{G.~Spadaccini}
\affiliation{Dipartimento di Scienze Fisiche, Universit\`a di Napoli "Federico II", I-80126, Napoli, Italy}
\affiliation{INFN, Sezione di Napoli, I-80126, Napoli, Italy}
\author{Sh.A.~Kalandarov}
\affiliation{Joint Institute for Nuclear Research, 141980 Dubna, Russia}
\affiliation{Institute of Nuclear Physics, 702132 Tashkent, Uzbekistan}
\author{C.~Beck}
 \affiliation{IPHC, IN2P3-CNRS, F-67037, Strasbourg Cedex2, France}
\author{S.~Barlini}
\affiliation{INFN, Sezione di Firenze, I-50125 Firenze, Italy}
\author{B.~Borderie}
 \affiliation{IPNO, IN2P3-CNRS and Universit\'e Paris-Sud 11, F-91406, Orsay Cedex, France}
\author{R.~Bougault}
\affiliation{LPC, IN2P3-CNRS, ENSICAEN and Universit\'e, F-14050, Caen Cedex, France}
\author{R.~Dayras}
\affiliation{CEA, IRFU, SPhN, CEA/Saclay, F-91191, Gif-sur-Yvette Cedex, France}
\author{G.~De Angelis}
\affiliation{INFN, LNL, I-35020 Legnaro (Padova) Italy}
\author{J.~De Sanctis}
\affiliation{INFN, Sezione di Bologna, I-40127 Bologna, Italy}
\author{V.L.~Kravchuk}
\affiliation{INFN, LNL, I-35020 Legnaro (Padova) Italy}
\author{P.~Lautesse}
\affiliation{IPNL, IN2P3-CNRS et Universit\'e, F-69622, Villeurbanne Cedex, France}
\author{N.~Le Neindre}
\affiliation{LPC, IN2P3-CNRS, ENSICAEN and Universit\'e, F-14050, Caen Cedex, France}
\author{J.~Moisan}
\affiliation {Grand Acc\'el\'erateur National d'Ions Lourds (GANIL), CEA/DSM-CNRS/IN2P3, Boulevard H. Becquerel, F-14076, Caen, France}
\affiliation{Laboratoire de Physique Nucl\'eaire, Universit\'e de Laval, Qu\'ebec, Canada}
\author{A.~D'Onofrio}
\affiliation{Dipartimento di Scienze Ambientali, Seconda Universit\`a di Napoli, I-81100, Caserta, Italy}
\author{M.~Parlog}
\affiliation{LPC, IN2P3-CNRS, ENSICAEN and Universit\'e, F-14050, Caen Cedex, France}
\author{D.~Pierroutsakou}
\affiliation{INFN, Sezione di Napoli, I-80126, Napoli, Italy}
\author{M.F.~Rivet}
\affiliation{IPNO, IN2P3-CNRS and Universit\'e Paris-Sud 11, F-91406, Orsay Cedex, France}
\author{M.~Romoli}
\affiliation{INFN, Sezione di Napoli, I-80126, Napoli, Italy}
\author{R.~Roy}
\affiliation{Laboratoire de Physique Nucl\'eaire, Universit\'e de Laval, Qu\'ebec, Canada}
\author {G.G.~Adamian} 
\affiliation{Joint Institute for Nuclear Research, 141980 Dubna, Russia}
\affiliation{Institute of Nuclear Physics, 702132 Tashkent, Uzbekistan}
\author {N.V.~Antonenko}
\affiliation{Joint Institute for Nuclear Research, 141980 Dubna, Russia}
\date{\today}

\pacs{24.60.Dr, 24.10.Pa, 25.70.Gh} 

\begin{abstract} 
Decay modes of excited nuclei are investigated in $^{78,82}$Kr + $^{40}$Ca reactions at 5.5 MeV/nucleon. Charged products were measured by means 
of the
$4\pi$ INDRA array. Kinetic-energy spectra and angular distributions of fragments with atomic number 
3 $\le Z \le$ 28 indicate a high degree of relaxation and are compatible with a fission-like phenomenon. Persistence of structure effects is evidenced from elemental cross-sections ($\sigma_{Z}$) as well as a strong odd-even-staggering (o-e-s) of the light-fragment yields. The magnitude of the staggering does not significantly depend on the neutron content of the emitting system. Fragment-particle coincidences suggest that the light partners in very asymmetric fission are emitted either cold or at excitation energies below the particle emission thresholds. The evaporation residue cross-section of the 
$^{78}$Kr + $^{40}$Ca reaction is slightly higher than the one measured in $^{82}$Kr + $^{40}$Ca reaction. The fission-like component is larger by $\sim$ 25\%  for the reaction having the lowest neutron-to-proton ratio. These experimental features are confronted to the predictions of theoretical models. 
The Hauser-Feshbach approach including the emission of fragments up to $Z$ = 14 in their ground states as well as excited states does not account for the main features of $\sigma_{Z}$. 
 For both reactions, the transition-state formalism reasonably reproduces the $Z$-distribution of the fragments with charge 12 $\le Z \le$ 28. However, this model strongly overestimates the light-fragment cross-sections and does not explain the o-e-s
of the yields for 6 $\le Z \le$ 10. The shape of the whole $Z$-distribution and the o-e-s 
of the light-fragment yields are satisfactorily reproduced within the dinuclear system framework which treats the competition between evaporation, fusion-fission and quasifission processes. The model suggests that heavy fragments come mainly from quasifission while light fragments are predominantly populated by fusion. An underestimation of the cross sections for 16 $\le Z \le$ 22 could signal a mechanism in addition to the capture process.

\end{abstract}

\maketitle
\section{Introduction}
 Heavy-ion induced reactions are appropriate to explore the response of nuclei under stress of different nature and to delineate the degrees of freedom at work in the various bombarding energy domains. 
 The regime of warm medium-mass ($A\sim100-130$) compound nuclei (CN) formed in fusion reactions at incident energies below 
10 MeV/nucleon is characterized by the predominant role of the angular momentum of the emitting nuclei and of the mass (charge) asymmetry degree of freedom. An abundant literature has reported that the CN decay modes populate the whole mass (charge) range from evaporated light particles (like  {\it n, p}, $\alpha$) up to the symmetric fission, and the intermediate-mass fragments (IMF) in between the two 
extremes~\cite{Mor1, Sobotka, Jing, Charity, Boger}. 
From the accumulated data one could identify two basic features of the final products:
the charge distribution evolves from a U-shape at low angular momentum (with a minimum at symmetry) towards a bell shape
 at high angular momentum (with a maximum around symmetric fission)~\cite{Sobotka}; a staggering of the fragment cross-sections $\sigma_{Z}$ is  superimposed on this global feature, with a magnitude which depends on the size of the emitting nuclei and which increases  as the neutron-to-proton $N/Z$ ratio of the emitter decreases ~\cite{Jing,Fan}. 
It has been suggested that the staggering effects reflect some properties of nuclei involved at the end of the disintegration cascade \cite{Ricciardi}. Indeed, a plausible explanation of the staggering
of
 $\sigma_{Z}$ would be  that structure effects persist in the production mechanism and that 
 fragments
  are emitted relatively cold, otherwise the subsequent decay would have blurred the fluctuations of the yields. 
 Moreover, the neutron content of the emitter manifests itself in the magnitude of the IMF cross-sections as shown in Refs.~\cite{Jing,Fan,Brzychczyk}. This raises the question of the $N/Z$  dependence of the decay channels  which is a relatively unknown and very attractive topic in the context of radioactive beam facilities.

 On the theoretical side, sophisticated approaches have been developed to explain the complex facets
 of the disintegration modes. Many features of the light-particle emission are satisfactorily understood within the Hauser-Feshbach  formalism~\cite{Hauser} emphasizing the role of the available phase space at each step of the whole cascade~\cite{Stokstad}. On the other hand, 
the mechanism at the origin of the fragment emission from CN has stimulated numerous approaches as for example: the multi-step Hauser-Feshbach model including emission channels up to Ca~\cite{Gomez88}; 
the transition~-state model describing IMF emission as asymmetric fission~\cite{Charity,Auger1987}; the dynamical cluster-decay model assuming pre-formed cluster~\cite{Gupta,Kumar}; the dinuclear system model aiming to treat the competition between the evaporation channel and the binary-decay channels associated to fusion and quasifission processes~\cite{Kalandarov}. 
Those approaches are based on distinct hypotheses as well as fundamental nuclear ingredients such as the level density or the fission barriers to describe the thermal and collective properties that rule the competition between CN decay modes. It is worth noticing that the $N/Z$  degree of freedom is expected to play a crucial role on these quantities. For example, the level-density parameter is related to the effective mass, a property of the effective nucleon-nucleon interaction that is sensitive to the neutron-proton composition of the nuclei; the fission barriers depend strongly on the symmetry energy that is weakly constrained by experimental data~\cite{Sierk}.
Therefore, new experimental data on decay channels of nuclei at high angular momenta and different $N/Z$ ratio are sorely needed. 

Besides the decay stage, the phase of CN formation has its own crucial interest. Indeed, since more than three decades, a rich wealth of data has revealed the complexity of the fusion process and of the collisional stage preceding the CN formation. For example, extensive experimental and theoretical investigations have shown that fusion mechanism at the vicinity of the barrier~\cite{LCorradi} is drastically influenced by the internal structure and $N/Z$ ratio of the participating nuclei. Moreover, a large body of data for a wide variety of systems has demonstrated the role of dynamical effects on the fusion process and the strong inhibition of the CN formation by quasifission (QF). This phenomenon corresponds to the separation of the partners after a significant rearrangement of the mass and charge degree of freedom~\cite{Bock,Back1985,Toke,Hinde2002,Hinde1999,Prasad2010}. Interestingly, in medium-mass systems, it has been recently shown~\cite{Kalandarov}, that the competition between fusion-fission and quasifission mechanisms strongly
 depends on the angular momentum. This calls for new data to extent our knowledge on the influence of the dynamics on fusion process in this mass region.

Finally, we would like to stress that an accurate prediction of the IMF 
cross-sections has important consequences. Indeed, one could perform spectroscopic
studies of the residual nuclei left in excited states
after the fragment emission. This kind of experiment has shown the strong 
selectivity of the $^{12}$C emission with respect to the 3$\alpha$ channel~\cite{Gomez98}. An evident area for such studies is around  the doubly magic $^{100}$Sn since these
nuclei are extremely difficult to reach by means of the standard fusion-evaporation method.
However, a recent attempt~\cite{La Commara} has suggested that the $^{12}$C emission from
a $^{116}$Ba CN formed in the $^{58}$Ni + $^{58}$Ni fusion reaction at $\sim$ 7 MeV/nucleon
does not offer a valuable alternative for producing a given isotope compared to the fusion-evaporation method. Therefore a better understanding of the IMF emission from medium-mass CN at low excitation energy is required.

For these reasons we investigated the decay modes of excited
nuclei produced in $^{78,82}$Kr + $^{40}$Ca reactions at 5.5 MeV/nucleon
incident energy. This energy regime is well adapted to form nuclei in a 
controlled way in terms of excitation energy since the incomplete-fusion process or pre-equilibrium emission are expected to be negligible. 
Exclusive measurements on an event-by-event basis are required to 
provide a characterization of the mechanism. Therefore a 4$\pi$
detection apparatus with low energy thresholds and charge identification of the products is needed. The combination of both INDRA array 
~\cite{Pouthas1995} and the technique of the reverse kinematics permit us to collect high quality data on evaporation-residues and elemental cross-sections of fragments. Our data set, obtained with a projectile pair differing by four neutrons,  gives new insights on the influence of the neutron content on decay mechanisms that allows us to evaluate the respective merits of very popular theoretical approaches. Some preliminary results have been recently presented~\cite{Bonnet2008}. Here we concentrate on main features of the heavy products, and the study of the light-particle emission will be presented in a forthcoming paper.

In Table~\ref{tab:tab0} are grouped some quantities characterizing the $^{78,82}$Kr + $^{40}$Ca reactions at 5.5 MeV/nucleon
incident  energy.   CN  excitation  energies   $E^{\star}$  have  been
calculated  using  mass   tables~\cite{Wap}.  $l_{graz}\hbar$  is  the
grazing   angular   momentum    given   by   semi-classical   formula.
$l_{pocket}\hbar$ is the  angular momentum at which the  pocket in the
interaction  potential vanishes.  The  potential is  calculated as  in
Ref.~\cite{Bass80}.  $J_{cr}\hbar$ is the maximum angular momentum for
capture process as deduced from the dinuclear system (DNS) calculations (see Sect.~V for details). $N/Z $ is the neutron-to-proton ratio of the reaction and $V_B$ is the fusion barrier~\cite{Bass80}. Others interaction potential choices, like those compared in~\cite{Dutt}, give similar $l_{pocket}$ and $V_B$ values. As reported in Table~\ref{tab:tab0}, 
the total available kinetic energy in the center-of-mass (c.m.) $E_{c.m.}$ is well above the fusion barrier and the grazing angular momentum is large with respect to $l_{pocket}\hbar$. Thus, in the reactions under study, we expect that the fusion process will be mainly governed by the inner pocket of the potential and to a lesser extent by the external fusion barrier.    
\begin{table}[htbp!]
\caption{\label{tab:tab0} Quantities characterizing the studied reactions.}

\centering
\begin{ruledtabular}
\begin{tabular}{l c c }
& \multirow{2}*{ $^{78}$Kr + $^{40}$Ca} &\multirow{2}*{$^{82}$Kr + $^{40}$Ca}\\
\\
 \hline
\\
 $E^{\star}$ (MeV)    & 99     & 107    \\
 $E_{c.m.}$/$V_B$      & 1.59   & 1.64   \\
$V_B$ (MeV)          & 91.2   & 90.3   \\
$N/Z$                  & 1.11   & 1.18   \\
 $l_{graz}$           & 96     & 100    \\
$l_{pocket}$         & 70     & 73     \\
 $J_{cr}$            & 73     & 75     \\
\\

 \end{tabular}
\end{ruledtabular}
\end{table}

The organization of the paper is as follows: the experimental procedures are described in Sec.~II. Experimental results are shown in Sec.~III for the inclusive data and in Sec.~IV for the fragment-light particle coincidences. Sec.~V deals with comparisons to statistical and dynamical calculations. The conclusions of this work are given in Sec.~VI.

\section{Experimental procedures}
\label{Expproc}
The experiment was performed at the GANIL facility in Caen.  
Beams  of $^{78,82}$Kr  projectiles with  energies of  5.5 MeV/nucleon
impinged on self-supporting 1 mg/cm$^{2}$ thick $^{40}$Ca targets. The
targets were prepared from high purity foils by rolling. The contaminants, mostly oxygen and tantalum, 
were negligible as thoroughly checked during the data analysis.

The kinetic energy and  atomic number of the ejectiles were measured by means of the 4$\pi$ INDRA array. 
The reverse kinematics confers to the reaction products a focussing at forward angles and a momentum boost in the laboratory frame. 
For the experimental data reported here, a significant part of the reaction products 
is emitted from 3$^\circ$ to 45$^\circ$. In this range, the INDRA array is made of 8 rings comprising detection modules with three layers: an  ionization chamber (IC) operated with 50 mbar (30 mbar)
of $\rm C_{3}F_{8}$ gas for 3$^\circ\le\theta_{lab}\le$~27$^\circ$ (27$^\circ\le\theta_{lab}\le$~45$^\circ$), respectively; a 300 $\mu$m thick silicon detector (Si); a 14 or 10 cm length CsI(Tl) scintillator. 
The medium and backward angles from 45$^\circ$ to 176$^\circ$ are covered by 8 rings equipped 
with IC/CsI(Tl) detectors, the ICs being operated with 30 mbar of  $\rm C_{3}F_{8}$ gas. For the calibration of the CsI at backward angles, one module per ring is equipped with a Si(80 $\mu$m)/SiLi(2000 $\mu$m) telescope inserted between IC and CsI. The energy calibration of the various layers  
was obtained by means of alpha particles emitted from a Cf source and from the 
elastic scattering of  projectiles having various energies 
($^{75,78,82}$Kr$^{12+}$, $^{75}$As$^{12+}$, $^{50}$Cr$^{12+}$, $^{100}$Mo$^{12+}$) selected thanks to the CIME \hbox{cyclotron}. Energy calibration of the detectors ensured on accuracy of 
within 5\%. 

The intensity of the beams was adjusted in order to maintain a low probability for pile-up of the events and the 
data acquisition dead time below 25\%. The reaction products were measured event-by-event by using two recording modes, a minimum-bias trigger based on the number $M$ of fired telescopes. The first mode ($M\ge$~1) ensures the measurement of the elastic scattering for normalization purposes while the second mode ($M\ge$~2) permits to accumulate high statistics for the reactions of interest. 
\begin{figure}[!t]
 \includegraphics*[width=8cm]{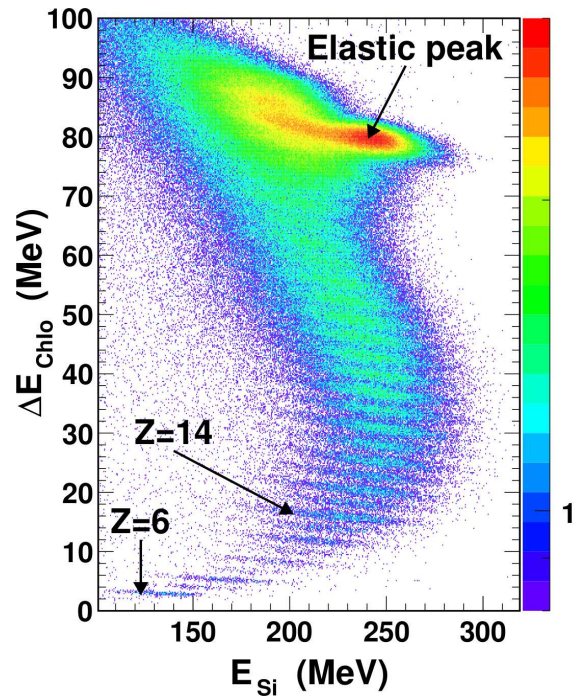} 
 \caption{(Color online) Two-dimensional plot combining the energy deposited in the ionization chamber (vertical axis) and in the silicon detector (horizontal axis) for fragments emitted at 10$^\circ\le\theta_{lab}\le~$14$^\circ$ measured in the $^{78}$Kr + $^{40}$Ca reaction at 5.5 MeV/nucleon.} 
 \label{fig.dee}
\end{figure}

The kinetic energy and the atomic number of the detected products were deduced from 
the energy deposited in the IC and Si detectors, corrected for the energy losses 
in the target as well as in the dead zones of the various detection layers~\cite{Kaliveda}. 
A typical example of a two-dimensional calculated spectrum used for the 
$Z$-identification is shown in Fig.~\ref{fig.dee} where the horizontal (vertical) 
axis represents the energy deposited in the Si (IC) detector, respectively. 
These data were obtained at  10$^\circ\le\theta_{lab}\le~$14$^\circ$. 
Although only the fragments emitted in the 
forward hemisphere in the c.m. are collected, one recognises the typical pattern of reaction products in reverse kinematics. 
The ridges associated to different atomic number 
are seen from $Z=6$ up to $Z=37$. The products with charge 3 $\le Z \le$ 5 
punched through the silicon detectors and they are 
identified by means of a two-dimensional plot (not shown here) built with the 
energies collected in the Si and CsI detectors. Interesting features could be extracted from these raw data. An odd-even-staggering is visible from 
the counting rates of the fragments up to $Z=16$ with a stronger
magnitude for fragments with charge $Z\le10$. 
Moreover, we clearly see a quasi-elastic component around $Z=36$ which manifests with a higher statistics. 

 Event-by-event $Z$-identification of each detected product was achieved by projecting data such as that of Fig.~\ref{fig.dee} onto lines which were drawn so as to follow the ridge for each $Z$. Charge resolution of one unit was obtained up to $Z=37$ for high-energy fragments. Identification for low-energy fragments was assured by calculations based on energy-loss tables, with a resolution of few charge units~\cite{Frankland}. Then we build two calculated spectra representing 
 the total kinetic energy  
in the laboratory frame $E_{tot}$ (the total charge $Z_{tot}$) 
obtained by summing up the kinetic energy (the atomic number) of 
each particle identified in the event, respectively. In the following steps of the analysis, 
we kept only the events satisfying $Z_{tot}\le$ 60 and  
$E_{tot}\le$ $E_{lab}$, where $E_{lab}$ is the bombarding energy. The limit on $Z_{tot}$ slightly exceeds the total available charge ($Z_{tot}=60$) to take into account the uncertainty on the charge identification. 
Applying such criteria enables us to control the event pile-up and double counting of the elastic scattering 
has been evaluated to be less than 4$\times10^{-6}$. 
Consequently, the number of events comprising particles coming from two distinct reactions is negligible.

\section{Experimental results}
\label{Expres}
\subsection{Kinematical features}
\begin{figure}[!t]
\includegraphics*[width=8cm]{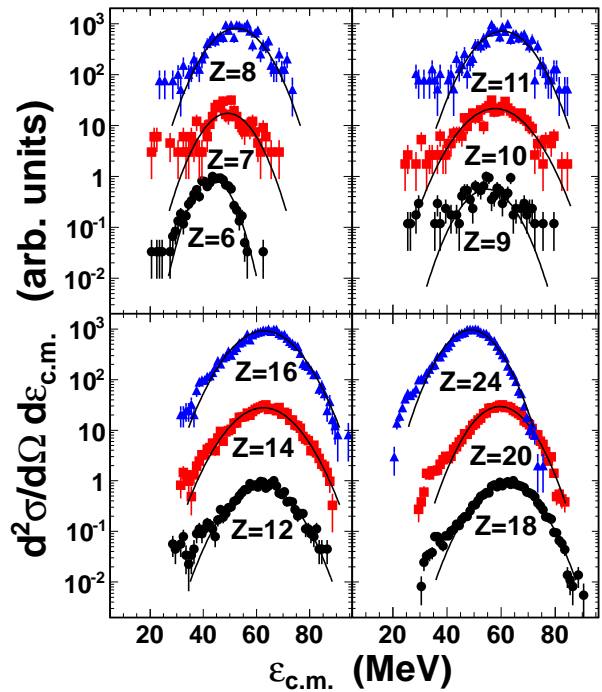} 
\caption { (Color online) Center-of-mass kinetic-energy spectra of fragments  with indicated atomic number from $Z=6$ to $Z=24$ produced in the $^{78}$Kr + $^{40}$Ca reaction at 5.5 MeV/nucleon and detected at 7$^\circ\le\theta_{lab}\le~$14$^\circ$. Lines represent the results of a fit  with a Gaussian function. Statistical errors are shown.} 
\label{fig.Ecmspectra}
\end{figure}

Another piece of  information on the reaction mechanism can be obtained from the kinetic-energy spectra of the ejectiles. 
The transformation into the center-of-mass frame was obtained by means of an event-by-event analysis. 
Fig.~\ref{fig.Ecmspectra} shows some representative examples of the c.m. kinetic-energy spectra of fragments with the indicated atomic number  from $Z=6$ to $Z=24$ scattered at 
7$^\circ\le\theta_{lab}\le~$14$^\circ$ in the $^{78}$Kr + $^{40}$Ca reaction 
at 5.5 MeV/nucleon. A Gaussian-like distribution (lines in Fig.~\ref{fig.Ecmspectra}) 
reproduces rather well the experimental data over a large energy range. Such a feature could be related to secondary emission of light particles or/and to shape fluctuations with the associated variations of the Coulomb barrier.

\begin{figure}
  \includegraphics*[width=8cm]{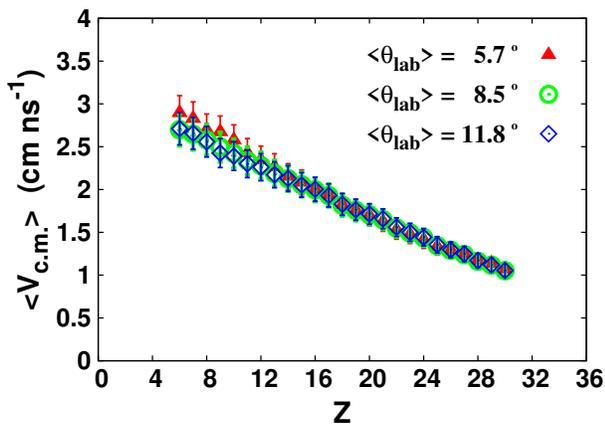} 
  \caption {(Color online) Experimental c.m. average velocity $<V_{c.m.}>$  of fragments with atomic number 6 $\le Z \le$ 28 measured at various angles in the $^{78}$Kr + $^{40}$Ca reaction at 5.5 MeV/nucleon.}
  \label{fig.velocity} 
\end{figure}

For each fragment, the c.m. average velocity $<V_{c.m.}>$ was deduced from the average kinetic 
energy assuming a mass number given by an empirical formula~\cite{Charity1988}. 
The results are reported in Fig.~\ref{fig.velocity} for various laboratory angles corresponding to the average values of the detection rings. 
For a given $Z$, $<V_{c.m.}>$ is roughly the same regardless of the emission 
angle except for $Z\le12$ at the most forward angles. We thus conclude that a high degree of  relaxation of the relative kinetic energy has been reached prior to the breakup of the excited 
nuclear system. $<V_{c.m.}>$ follows a quasi-linear decreases with increasing atomic charge $Z$. 
This feature is well documented~(\cite{Jing,Charity,Auger1987}), 
and is interpreted as a signature of a binary process 
dominated by the Coulomb interaction between the considered fragment and its complementary partner. 
The total average kinetic energy for symmetric division ($<TKE_{sym}>$ = 81~$\pm$~2 MeV for $Z=28$) 
is consistent ($E_{K}$ = 83.4 MeV for the $^{118}$Ba nucleus) with a recent compilation  on the total kinetic energy release 
in the fission phenomenon~\cite {Beck1989}. 

\subsection{Angular distributions}
\begin{figure}[!t]
  \includegraphics*[width=8cm]{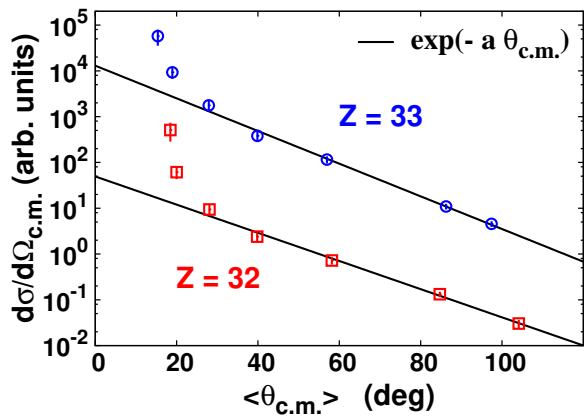} 
\caption {(Color online) Angular distributions of fragments  with atomic number $Z=32$ and 33 
produced in the $^{78}$Kr~+~$^{40}$Ca reaction at 5.5 MeV/nucleon. The lines are exponential functions to guide the eye.}
\label{fig.angulardist1}
\end{figure}

\begin{figure}[!t]
  \includegraphics*[width=8cm]{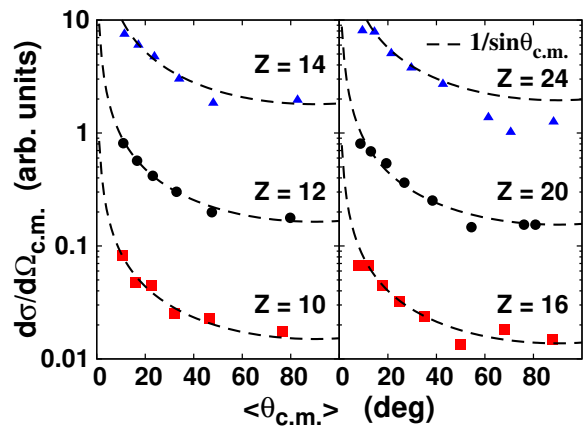} 
\caption {(Color online) Angular distributions of fragments  with charge $Z=10, 12, 14, 16, 20, 24$ produced in the $^{78}$Kr~+~$^{40}$Ca reaction at 5.5 MeV/nucleon. 
Dashed lines are $1/\sin\theta_{\rm c.m.}$ functions that have been normalized to the experimental data at $<\theta_{lab}>$ = 5.7$^\circ$, corresponding to $<\theta_{c.m.}>$~=~12$^\circ$--17$^\circ$. Error bars are inside the symbols.}
\label{fig.angulardist2}
\end{figure}

Valuable information on the production mechanism could be extracted from the angular distributions of the fragments. These distributions are obtained by integrating the 
kinetic-energy spectra. Some typical examples are given in 
Figs.~\ref{fig.angulardist1} and~\ref{fig.angulardist2} 
for various fragments. 

The angular distributions of the fragments with atomic number close to the projectile 
one ($Z=36$) are strongly peaked at forward angles as shown in Fig.~\ref{fig.angulardist1}. These products arise from direct two-body 
reactions or deep inelastic collisions  in which nucleons are transferred into or emitted 
from the projectile. Indeed, in peripheral collisions the target-like products are expected to be ejected in the backward hemisphere of the c.m., while those coming 
from the projectile would be strongly focused in the forward hemisphere. 
Fig.~\ref{fig.angulardist1} 
illustrates such a behaviour for $Z=32$ and $Z=33$ for which the angular distributions 
$d\sigma/d\Omega_{\rm c.m.}$ exhibit a strong decrease. Moreover, one observes two components corresponding presumably to quasi-elastic reactions at the most forward angles and deep-inelastic collisions which dominate for 
$\theta_{c.m.}\gtrsim~$20$^\circ$. The continuous line in Fig.~\ref{fig.angulardist1} represents an exponential function that follows the experimental data for $\theta_{c.m.}\gtrsim~$20$^\circ$. 
 
In Fig.~\ref{fig.angulardist2} we present the angular distributions 
$d\sigma$/d$\Omega_{\rm c.m.}$ for fragments with atomic number $Z=10, 12, 14, 16, 20, 24$
 produced in the $^{78}$Kr + $^{40}$Ca reaction. In spite of a measurement over a limited angular range in the 
laboratory frame, the reverse kinematics allows to 
define unambiguously the shape of the angular distributions in 
the c.m. frame. In contrast with the previously observed feature for fragments with 
$Z\sim$~36, the angular distributions  follow a $1/\sin\theta_{\rm c.m.}$ 
dependence (shown as dashed lines in Fig.~\ref{fig.angulardist2}). 
This signs a 
high degree of equilibration. Indeed, in heavy-ion reactions, CN which undergo fission have generally high angular momentum and the angular distributions of the fission fragments would show a $1/\sin\theta_{\rm c.m.}$ shape. However, this kind of behaviour is not a sufficient condition to ensure a CN formation. In fact, in quasifission (QF) process, the reactants retain some memory of the entrance channel which manifests in a strong anisotropy of the angular distribution~\cite{Hinde1999}. Our apparatus does not allow an accurate measurement of the angular distributions of the fragments scattered at angles close to the beam direction. This prevents a dedicated investigation of the anisotropy. Thus at this stage of the analysis of the angular distributions presented in Fig.~\ref{fig.angulardist2}, one concludes that the 
predominant mode of the fragment production is the disintegration 
either of a long-lived system or CN.

\subsection{Fragment-fragment coincidences}
\begin{figure}[!t]
\includegraphics*[width=7cm]{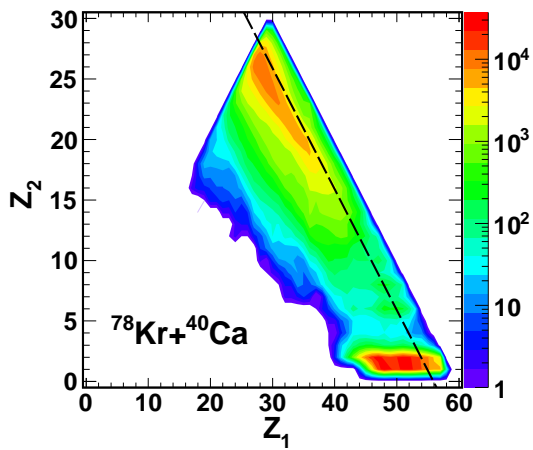} 
\caption {(Color online) Experimental correlation between the two biggest fragments $Z_{1}$ and $Z_{2}$ with 
 $Z_{1}\ge Z_{2}$  and 48~$\le~Z_{tot}~\le$~60.}
\label{fig.Z1Z2}
\end{figure}
The correlations between the charge of the fragments are instructive since they permit 
to check the binary nature of the mechanism. In the present work,  
an even-by-event analysis was performed in order to extract the two biggest fragments, $\it{i.e.}$ those having the highest atomic numbers $Z_{1}$ and $Z_{2}$ (with $Z_{1}$~$\ge$~$Z_{2}$) in each event.  Fig.~\ref{fig.Z1Z2} 
shows the $Z_{1}$ {\it vs} $Z_{2}$ correlation measured in the $^{78}$Kr~+~$^{40}$Ca reaction 
in the case of events satisfying the criterion 48~$\le Z_{tot}\le$~60. The lower limit is 
applied to exclude the events in which one of the two fragments has not been detected. The upper limit take into account the uncertainty on the $Z$-identification (see Sec.~\ref{Expproc}). The highest yields are localised in two regions:  
$Z_{1}\sim$~50 and  $Z_{2}\sim$~2  corresponding to the evaporation channel in one side; the region with 
$Z_{1},Z_{2}~\sim$~25--30 representing the symmetric fragmentation mode in another side. 
The residues exclusively populated after light-particle emission could be well separated from those populated by IMF emission. This is important to underline since in case of a competition between CN and QF processes, one could unambiguously associate evaporation residues (ERs) with CN formation. 
The ridge of the counting rates seen in Fig.~\ref{fig.Z1Z2}  is slightly shifted to an average value smaller by about two charge units than the total available 
charge 
($Z=56$), reflecting the light-particle emission from the fragments, or/and from the composite system before splitting. 
The linear correlation between $Z_{1}$ and $Z_{2}$ 
illustrates the binary nature of the mechanism. Here, the term binary means that 
the major part of the nucleons available in the reaction is distributed in the two biggest measured fragments.

As far as kinetic-energy spectra, angular distributions of the fragments and fragment-fragment coincidences are concerned, the same conclusions hold for $^{82}$Kr + $^{40}$Ca \hbox{reaction}.

\subsection{Cross sections}
The absolute differential cross-sections $d\sigma/d\Omega_{\rm c.m.}$ were obtained from the normalization with respect to the elastic scattering. 
To select the appropriate angle for normalization purposes, 
both grazing angle and angular distribution of the elastic scattering 
were deduced from optical model calculations~\cite {Fresco}.  
To do so, a set of optical parameters was extracted from the study of 
the Ar~+~Se reaction at 5 MeV/nucleon~\cite{ArSe} which is similar to those studied in the present work. From the analysis, we deduced that the grazing angle is about $\theta_{lab}$~=~20$^\circ$ (around $\theta_{c.m.}$~=~55$^\circ$). 
Moreover,  $\sigma/\sigma_{Ruth}(\theta_{lab})$~=~1 for $\theta_{lab}\le~$14$^\circ$. 
Thus the Rutherford differential cross-section of the elastic scattering was 
integrated over the range 7$^\circ\le\theta_{lab}\le~$10$^\circ$ 
to get the normalization factor. Then the absolute total cross-sections of the fragments with  atomic number 3~$\le Z \le$~28 
were obtained by angular integration assuming a $1/\sin\theta_{\rm c.m.}$
shape as indicated in Sec.~III.B. This procedure could not be suited to the non-measured part of the angular distribution close to the beam direction, but the weight of this angular domain is negligible. 

In the following, we concentrate on the decay behaviour of a long-lived system, 
and consequently the cross sections of the quasi-elastic component are not discussed here due to 
the exponential shape of the angular distributions, akin to a fast process.

\begin{figure}[!b]
  \includegraphics*[scale=0.6]{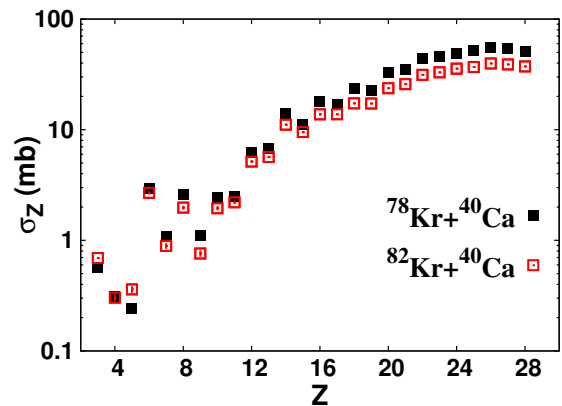} \\
\caption{(Color online) Experimental cross-sections for fragments with atomic number 3 $\le Z \le$ 28 emitted in the $^{78}$Kr~+~$^{40}$Ca (full squares) and $^{82}$Kr~+~$^{40}$Ca (open squares) reactions at 5.5 MeV/nucleon.}
\label{fig.CSexp}
\end{figure}

 The inclusive cross-sections $\sigma_{Z}$ of fragments with atomic number 3 $\le Z \le$ 28 
 are shown in Fig.~\ref{fig.CSexp} 
 for the $^{78}$Kr + $^{40}$Ca (solid squares) and $^{82}$Kr + $^{40}$Ca (open squares) reactions. Note that the Be cross-sections are depleted due to the contribution of the non-identified $^{8}$Be fragment. The $\sigma_{Z}$ distributions for both systems exhibit a maximum around $Z=26$, a value close 
 to half of the available charge. Such a feature indicates that these elements come either from 
 the symmetric fission of CN or from a class of collisions in which 
 a strong relaxation of the entrance channel mass-asymmetry has been reached. 
 Moreover, except for 3 $\le Z \le$ 5,  $\sigma_{Z}$ measured in the $^{82}$Kr + $^{40}$Ca system is systematically lower 
 and the yields around the symmetric splitting are about 25\% smaller for the system having the highest neutron-to-proton ratio. Such a lowering of the cross section for the symmetric splitting as 
 the neutron content of the emitter increases is also observed in 
 $^{78,82,86}$Kr + $^{12}$C reactions~\cite{Jing}. 
 This $N/Z$ dependence would be consistent with the expectations of the 
 liquid-drop model in which the fission barrier of a neutron-poor CN is 
 expected to be smaller than for the neutron-rich one, providing that these fission-like fragments originate from CN decay.

A strong odd-even-staggering (o-e-s) of the $\sigma_{Z}$ for fragments with $Z\le10$ 
is visible, and this effect is still present for higher $Z$ with a smaller amplitude. 
Fluctuations in fragment yields have already been observed in a wide range of 
reactions, from CN 
regime to spallation reactions~\cite{Jing,Fan,Ricciardi,Steinhauser,Cavallaro}. It is worth noticing that the staggering in the 
yields of light clusters shown in Fig.~\ref{fig.CSexp} is very similar to the 
one observed for systems in the same range of mass, excitation energy and 
angular momentum~\cite{Sobotka,Boger}. This  would indicate that the 
staggering is not preferentially driven by microscopic properties of the  
complementary partners since they are different for each studied reaction. 

As shown in Fig.~\ref{fig.CSexp}, the o-e-s for light fragments  is roughly the same for both reactions and is about a factor 3. Such a result is at variance with 
$^{78,82,86}$Kr + $^{12}$C data~\cite{Jing} for which the o-e-s decreases for neutron-rich CN. As far as the entrance channels are concerned, the main difference between those data and the present ones  comes from the magnitude of the spin that could be transferred into the composite system. Thus, the o-e-s of the light-fragment yields could be influenced by the spin which would induce different 
compactness of the scission-configurations and thus a sensitivity to structure properties in the deformation space. 

As suggested by the shape of the $Z$-distribution, the high partial waves in the entrance channel should have fed the fragment emission mechanism. However, the cross sections of the light clusters (Li, B) are astonishingly low. Indeed, in  
$^{93}$Nb + $^{9}$Be, $^{12}$C reactions~\cite{Charity} in which low angular momentum were involved, the cross sections of the light clusters are of the same order of magnitude or even higher than in our measurements. A possible explanation would be that at high angular momentum a large part of the flux  has been deviated from a CN formation. Such a possibility will be discussed in Sec.~V. 

The cross sections  of the fission-like products, $\sigma_{fiss}^{exp}$, were obtained by summing up the yields of the fragments 
in a range of  atomic number 3 $\le Z \le$ 26. 
The upper limit corresponds to the atomic number of the fragments produced with the highest 
cross-section and takes into account qualitatively the secondary decay of light charged particles (see 
Fig.~\ref{fig.Z1Z2}). 
Thus, considering the range 3 $\le Z \le$ 26 
we obtain  $\sigma_{fiss}^ {exp} = 447\pm$~46 mb ($\sigma_{fiss}^{exp} = 332 \pm~35$ mb) for the  $^{78}$Kr + $^{40}$Ca ($^{82}$Kr~+~$^{40}$Ca) reactions, respectively. We remind here that we have termed as fission-like products those with an angular distribution akin to that of a long-lived system, and $\sigma_{fiss}^{exp}$ could contain both CN and QF contributions. 

The ER component is identified thanks to a  $\Delta E-E$ 
two-dimensional plot using the energy deposited in the IC and Si detectors. 
Absolute differential cross-sections $d\sigma_{ER}/d\Omega_{lab}$ are deduced 
from the normalization  with respect to the elastic scattering. Since 
$d\sigma_{ER}/d\Omega_{lab}\approx\exp[-k\sin^{2}\theta_{lab}]$~\cite{Gomez79}, 
the experimental distribution is extrapolated towards the beam direction, 
and $\sigma_{ER}^{exp}$ could be extracted. Extensive simulations using statistical 
code PACE4~\cite{Tarasov2005} were performed to check this procedure. 
We  obtain $\sigma_{ER}^{exp}$ = 539 $\pm$~100~mb ($\sigma_{ER}^{exp}$ = 492 $\pm$~90~mb) 
for the $^{78}$Kr + $^{40}$Ca ($^{82}$Kr + $^{40}$Ca) reactions, respectively. These cross sections together with $\sigma_{fiss}^{exp}$ are gathered in Table~\ref{tab:tab1}.

 The sum of the fission-like and ER cross-sections defines the experimental capture cross-sections $\sigma_{capt}^{exp}=\sigma_{ER}^{exp}+\sigma_{fiss}^{exp}$ and we measured  $\sigma_{capt}^{exp}$ = 986 $\pm$ 110~mb  ($\sigma_{capt}^{exp}$ = 824 $\pm$ 97~mb)    
 for the  $^{78}$Kr + $^{40}$Ca ($^{82}$Kr~+~$^{40}$Ca) reaction, respectively.  
 By using the sharp cut-off approximation formula
\begin{eqnarray}
\nonumber\sigma_{capt}^{exp}(E_{\rm c.m.})&=&\frac{\pi\hbar^2}{2\mu E_{\rm c.m.}}\sum_{J=0}^{J_{max}}(2J+1)\\
&=&\frac{\pi\hbar^2}{2\mu E_{\rm c.m.}}(J_{max}+1)^2,
\label{sigfus}
\end{eqnarray}
we obtained $J_{max}^{exp}$ = 75 $\pm$ 4 (70 $\pm$ 4) for the $^{78}$Kr + $^{40}$Ca
($^{82}$Kr + $^{40}$Ca) reaction, respectively. 
\begin{table}[htbp!]
\caption{\label{tab:tab1} Measured and calculated evaporation residues and  fission-like cross-sections. See Sec.~V. for details of the calculations performed with GEMINI and DNS codes.}
\centering
\begin{ruledtabular}
\begin{tabular}{l  c  c }
\multicolumn{1}{c}{\multirow{2}*{(mb)}} & \multirow{2}*{  $^{78}$Kr~+~$^{40}$Ca   } & \multirow{2}*{  $^{82}$Kr~+~$^{40}$Ca   }\\
\\
\hline
\\
$\sigma^{exp}_{fiss}$    & 447 $\pm$ 46  & 332 $\pm$ 35 \\[0.5em]
$\sigma^{exp}_{E.R.}$    & 539 $\pm$ 100 & 492 $\pm$ 90 \\[0.5em]
$\sigma^{gemini}_{fiss}$ & 600         & 547         \\[0.5em]
$\sigma^{gemini}_{E.R.}$ & 237         & 285         \\[0.5em]

$\sigma^{DNS}_{fiss}$    & 349             & 208        \\[0.5em]
$\sigma^{DNS}_{E.R.}$    & 601             & 638         \\
\\
\end{tabular}
\end{ruledtabular}
\end{table}

From the ER cross-sections we have calculated the reduced quantity 
$\Lambda_{ER}= 2\mu E_{\rm c.m.}\sigma /(\pi\hbar^2)$, in
 which the dependence on the entrance channel is removed.
In the literature we have extracted the same quantity for reactions similar to those studied here. The $\Lambda_{ER}$ values for
$^{78,82}$Kr~+~ $^{40}$Ca reactions are compatible with the data for 
quasi-symmetric entrance channel such as, for example, $^{58}$Ni + $^{64}$Ni~\cite{Rhem94} or
$^{52}$Cr + $^{56}$Fe~\cite{Agarwal80} and mass-asymmetric as $^{32}$S + $^{76}$Ge~\cite{Guillaume} reaction. However the $\Lambda_{ER}$ values for
$^{78,82}$Kr + $^{40}$Ca reactions are smaller
than the one extracted for other mass-asymmetric systems such as $^{16}$O + $^{92}$Mo~\cite{Agarwal80}
and $^{32}$S + $^{100}$Mo~\cite{Brondi}. This would indicate a different boundary between
evaporation and fission-like channels in the $J$-space as a function of the mass-asymmetry of the entrance channel, as for example when fusion and quasifission processes compete with each other.

The capture cross-section in $^{78}$Kr + $^{40}$Ca reaction is higher than the one measured in $^{82}$Kr + $^{40}$Ca reaction. This behaviour is at variance with observations in the vicinity of the Coulomb barrier for systems with similar masses~(\cite{Stefanini0,Stefanini1,Zhang}). Considering these measurements at 
 the highest bombarding energy ($\sim10\% $ above the Coulomb barrier),  $\sigma_{capt}^{exp}$ of a neutron-rich system ($^{36}$S~+~$^{96}$Zr) exceeds by  $ \sim25\%$ the capture cross-section of a neutron-poor system ($^{36}$S~+~$^{90}$Zr) and the same trend is observed for the $^{32}$S~+~$^{90,96}$Zr reactions. However, in these cases the cross sections of fission-like products were negligible while this decay mode accounts for almost $50\%$ of $\sigma_{capt}^{exp}$ in the $^{78,82}$Kr + $^{40}$Ca reactions at 5.5 MeV/nucleon.  
In the reactions studied here,  the difference in $\sigma_{capt}^{exp}$ is mainly due to the fission-like 
component, leading to a smaller capture cross-section for the $^{82}$Kr + $^{40}$Ca system. The confrontation with the predictions of theoretical models will bring more information to discuss this aspect.
  
\section{Fragment-particle coincidence measurements}
To better understand the fragment  emission mechanism and to get more insights on the o-e-s of the light-fragment yields, we  have performed an
event-by-event  analysis of  the  light charged particles (LCPs)  in
coincidence with fragments. In the first step, we calculated for each fragment
the relative velocity between that  fragment and each detected LCP of the  event. Then we consider a new frame with one axis corresponding to the direction  of the fragment velocity in  the c.m. frame and the plane perpendicular to this axis. Finally, we projected the relative velocities previously calculated onto this new frame and deduced  the
component  parallel  ($V_{\parallel}$)  and  perpendicular
($V_{\perp}$) with respect to the direction  of the fragment velocity in  the c.m. frame. In doing so, for fragments of a given $Z$, having different emission angles in the c.m., the procedure enables to
construct a common reference frame for the LCPs in coincidence with these fragments. We have seen the binary nature of the fragment production with a small amount of particles emitted meanwhile. Thus, the kick induced by the emitted particles should be small and one could assume that fragments are flying back-to-back in the center-of-mass. Then, the emission direction of one fragment defines the recoil direction of its complementary partner. With such a method applied to an ensemble of reactions, the particles emitted by one fragment with a constant velocity value will draw one circle centered at the origin of the reference frame in a V$_{\parallel}$-V$_{\perp}$ plot.

\begin{figure}[!b]
  \centering
  \includegraphics*[width=7cm]{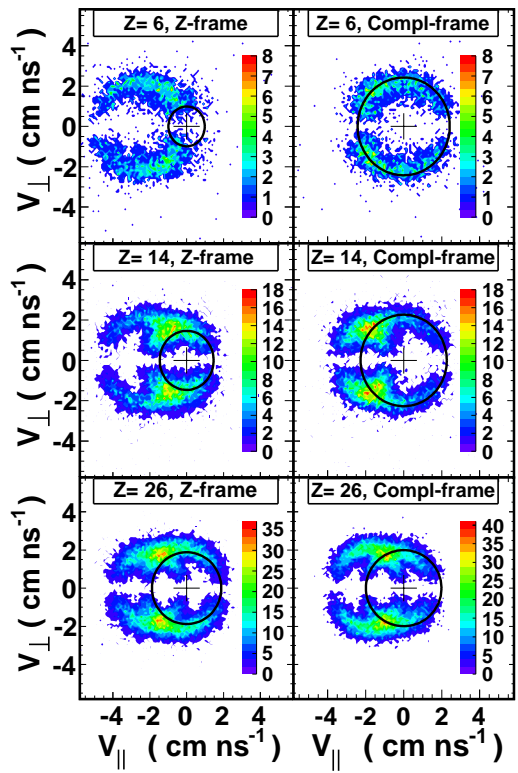}\\
  \caption{(Color online) V$_{\parallel}$-V$_{\perp}$  diagrams  of  alpha  particles
    detected in  coincidence with  C (first row), Si (second row)  and Fe  (third row)
    fragments produced in $^{78}$Kr + $^{40}$Ca reaction at 5.5 MeV/nucleon (see text). The
    velocities  are calculated in  the reference  frames of  the light
    fragment  (left  panels)  and of  the complementary  fragment  (right
    panels) }
  \label{fig.LCPFragcoinc}
\end{figure}

Fig.~\ref{fig.LCPFragcoinc} presents typical examples of
$V_{\parallel}-V_{\perp}$ diagrams  for  $\alpha$-C (first row), $\alpha$-Si (second row) and  
$\alpha$-Fe (third row) coincidences measured in the $^{78}$Kr~+~$^{40}$Ca reaction . The  black circles represent
the average velocities  taken from systematics compiled by  Parker {\it et al.}~\cite{Parker1991}. For $\alpha$-C coincidences, the relative velocities draw a circular region (akin of a Coulomb ring) which is centered at
the origin when they are projected into the frame (termed as Compl-frame) of the complementary partner of the C nuclei  (top right panel) whereas no such a circular region centered at the origin can be seen when the relative velocities are plotted in the frame (termed as Z-frame) of the light partner (top  left panel). For $Z=14$ and 26,
both  fragments emit light-particles as illustrated by  the two circles  centered  at  both  reference  frames. 
Thus,  we  observe  the change of behaviour of the light-particle emission from very asymmetric 
($Z=6)$ to asymmetric ($Z=14$) and almost symmetric ($Z=26$) fragmentation. The same
conclusions  hold  for  fragment-proton coincidences.  Thus, in  $^{78}$Kr~+~$^{40}$Ca reactions at  5.5 MeV/nucleon, the  LCPs are
emitted by  both fragments in the case  of symmetric fragmentation,  while for a
very asymmetric fragmentation, only  the heavy fragment emits particles. The
main lesson to be learnt is that the  light fragments are either produced cold
or at excitation energies below the proton or alpha emission thresholds. Extensive simulations were performed in
order  to check  that these  results  are not  related to  the geometrical
acceptance since the present analysis has been performed with fragments and particles detected at 3$^\circ\le\theta_{lab}\le~$70$^\circ$. Such a limited angular range prevents to extract quantitative information on emission characteristics such as multiplicity of light-charged particles associated to each fragment pair. This kind of analysis will be presented in a forthcoming paper. 

\begin{figure}
  \centering
  \includegraphics*[width=8cm,angle=0]{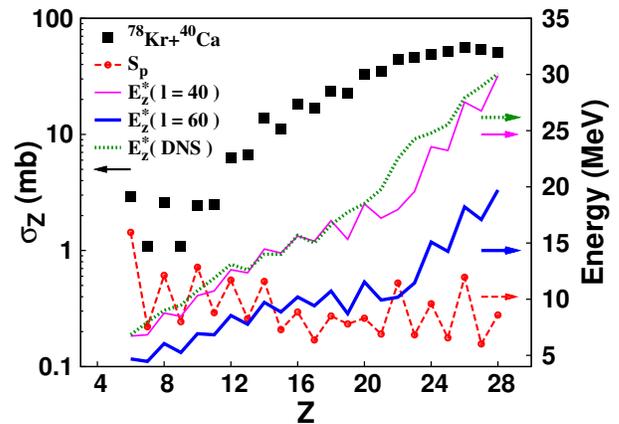}\\
  \caption{(Color online) Experimental  cross-sections for  fragments  emitted  in
    $^{78}$Kr~+~$^{40}$Ca (solid squares) reactions at 5.5 MeV/nucleon. 
    The  broken dashed line  represents the  proton
    separation energy. Thin (thick) lines refer to the excitation 
    energy stored in the fragment assuming an initial spin of 40 (60) respectively. Dotted line shows the DNS calculations. }
  \label{fig.EsepExp}
\end{figure}

The broken dashed line in Fig.~\ref{fig.EsepExp} shows the proton separation energy
$S_{p}$ calculated  for the  most abundant element  given by  the mass
tables. A strong
o-e-s is observed for $S_{p}$ with roughly the same
magnitude over the range 6~$\le~Z~\le$~28. It is worth noticing that the
o-e-s of $S_{p}$ and $\sigma_{Z}$ are in phase each other. For light fragments both
$\sigma_{Z}$ and $S_{p}$ are larger  for even-$Z$. 
One can make an estimation of the excitation energy $E_{Z}^{*}$ stored in the fragments. The total kinetic energy released in the binary fragmentation could be deduced from the kinetic
energy of the light partner for which the mass number is calculated assuming that its
$N/Z$ ratio is the same as the composite system. By assuming a rigid rotation
and a thermal equilibrium between both partners one can deduce $E_{Z}^{*}$ from the energy balance.
The results of such calculations are shown in Fig.~\ref{fig.EsepExp}
for an initial angular momentum of 40 (thin line) and  60 (thick line). $E_{Z}^{*}$ increases almost constantly from about 8 MeV for $Z=8$ to about 30 MeV for $Z=28$. The staggering
 of $E_{Z}^{*}$ is due to the fact that isotopic distribution for a given $Z$ is not taken into account.  
The values of $E_{Z}^{*}$ for $Z\le$~12 are below 15 MeV, $\it{i.e.}$ do not exceed the separation energy.
One should note that the particle-fragment Coulomb barrier is not included, as it would have been done to estimate the emission energy thresholds. However, taking into account the Coulomb barrier would not change drastically the pattern since the Coulomb energy grows smoothly with the atomic number of the fragment.  

The attenuation of
the  staggering of $\sigma_{Z}$  for fragments having large $Z$ would  be related  to a
blurring due  to light-particle emission as  suggested by the
coincidence data and by the estimation of $E^{*}$ for symmetric fragmentation. Same conclusions hold when considering the separation energy of alpha particles.
Thus,  the $\sigma_{Z}$ for light fragments reflect the
persistence of structure effects in asymmetric fragmentation. This could be
associated to a microscopic contribution to the potential energy surface which
is a  key ingredient  in determining the fragment yields and/or to specific
properties of the level density at energy below the particle emission thresholds.
Such influences need further investigations.

\section{Comparison with models}
In this section we compared data and the predictions of three theoretical approaches: two of them describe the decay modes of CN while the third one treats the dynamical evolution of the interacting partners and the physics governing the CN formation. Comparison of preliminary data and the dynamical cluster-decay model assuming pre-formed clusters~\cite{Gupta} has been presented in Ref~\cite{Kumar}.
\subsection{Comparison with BUSCO}
The Hauser-Feschbach approach is very successful in computing the light-particle emission from CN. In the BUSCO code~\cite{Gomez88}, this formalism has been extended to the IMF emission in their ground states as well as excited states. In the version of the code we used in the present work, the emission of fragments up to $Z=14$ has been incorporated. It should be noticed that fission channel is not taken into account. However, the model contains interesting features which justify the comparison to the present data, providing that the CN spin-distribution is given by the sharp cut-off approximation with $J_{max}$ kept as a free parameter.

The decay width of a channel $\alpha$ from a CN formed at a spin
$J$ is given by~\cite{Stokstad,Gomez88}
\begin{equation}
P_{\alpha}^{J}=\sum_{l_\alpha}\int {T_l}_{\alpha}
( {\epsilon}_{\alpha})\rho ({E^*_{CN}-\epsilon}_{\alpha},J)d{\epsilon}_{\alpha}.
\label{PJ}
\end{equation}
In Eq.~\ref{PJ}, $\it {T_l}_{\alpha}$ are the optical-model transmission coefficients
evaluated at the relative kinetic energy ${\epsilon}_{\alpha}$ in the emitter frame and
$\rho$ is the Fermi-gas model level density of the daughter nuclei computed with the prescription of
Ref.~\cite{Cameron}. The transmission coefficients have been parameterized by a Fermi function
\[
 {T_l}_{\alpha}( {\epsilon}_{\alpha})=(1+\exp[-({B_l}_{\alpha}-{\epsilon}_{\alpha})/{\Delta}_{\alpha}{B_l}_{\alpha}])^{-1},
\]
where
\[
 {B_l}_{\alpha}= \it B_{0}+ {\hbar^2} l_{\alpha} (l_{\alpha}+1) / {2\mu R_{\alpha}^2}.
\]
The parameters $B_{0}$,  $R_{\alpha}$ and ${\Delta}_{\alpha}$   are obtained
from the best fits of optical-model transmission coefficients. The predictions of the model have been successfully compared to data in medium-mass CN region~\cite{Brzychczyk,Gomez88,Gomez98}.

\begin{figure}
  \includegraphics*[scale=0.65]{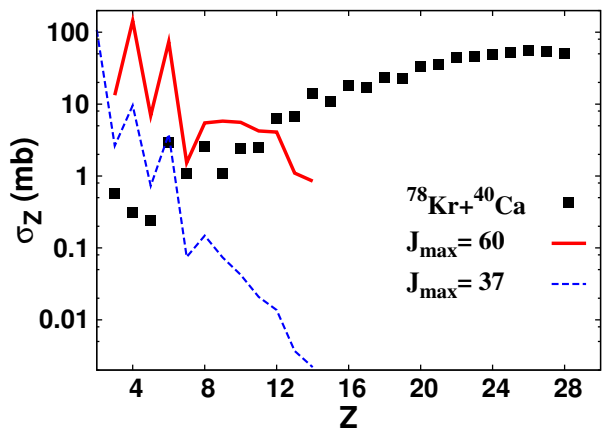}
  \caption{(Color online) Experimental cross-sections for fragments emitted in the $^{78}$Kr + $^{40}$Ca reaction at 5.5 MeV/nucleon (squares), compared to BUSCO calculations assuming a $J$-distribution given by the sharp cut-off approximation with $J_{max}=60$ (thick line)  and $J_{max}=37$ (dashed line). Calculations have been performed with a level-density parameter $a=A/8.5$ MeV$^{-1}$.}
\label{fig.CSBUSCO}
\end{figure}

The present calculations were performed using a level-density parameter $a=A/8.5$ MeV$^{-1}$ and
a sharp cut-off approximation with $J_{max}$~=~60 as a starting guess. The results of the BUSCO calculations for the $^{78}$Kr~+~$^{40}$Ca reaction are symbolized by a thick line in Fig.~\ref{fig.CSBUSCO}.
The model fails to reproduce the features of the $Z$-distribution, although an odd-even staggering as in the data is seen for $Z\le$~8. For $Z\le$~14  one observes
a global decreasing of the calculated $\sigma_{Z}$ at variance with data.
More specifically, the cross section of C is overestimated by a factor 30,
while $\sigma_{Z}$ for $8\leq~Z~\leq$12 are overestimated within a factor of 2 to 6.
A calculation assuming $J_{max}$= 37 (dashed line in Fig.~\ref{fig.CSBUSCO}) in order to reproduce $\sigma_{Z}$ for C largely misses the yields of the other species. Taking a $J$-distribution with a diffuseness
around $J_{max}$ instead of a sharp cut-off approximation, or making different choices of the level-density parameter do not improve the predictions of the model. 

 \begin{figure}[!b]
   \includegraphics*[scale=0.4]{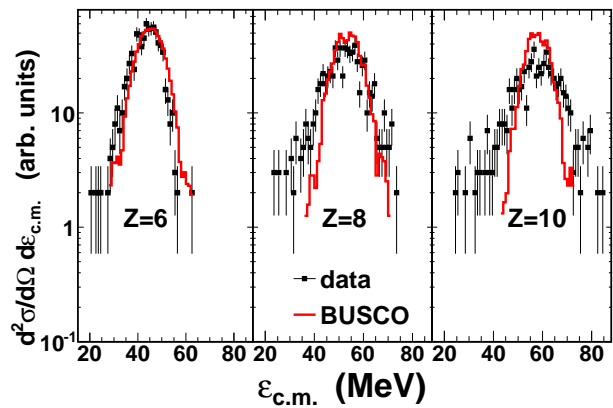}
  \caption {(Color online) Kinetic energy spectra in c.m. frame for $Z=6, 8, 10$ emitted in the 
  $^{78}$Kr + $^{40}$Ca reaction. Histograms are data and dashed lines are the results from the BUSCO calculations using $J_{max}=60$ and a level-density parameter $a=A/8.5$ MeV$^{-1}$. Calculations were normalized to data assuming the same integral for each $Z$.}
\label{fig.BUSCOSpectra}
\end{figure}
Since the interaction barriers play a crucial role in the competition between the decay channels, we compared the calculated kinetic-energy spectra of the fragments to the experimental data. In the BUSCO code, the kinetic-energy spectra result from the folding of the optical-model transmission coefficients and the level density. Thus the shape of the spectra is a good test of the calculation. The comparison of theoretical and measured spectra is presented in Fig.~\ref{fig.BUSCOSpectra} for $Z=6, 8, 10$. For each $Z$, the calculation was normalized to the integral of the kinetic energy distribution. The agreement is very good for the mean kinetic energy. However, the calculated  width of the distribution is smaller. The same conclusion holds for other
fragments. Improvement of the calculated kinetic-energy spectra could be obtained by a fine-tuning of the parameters, but the isotopic distribution is unknown and such a fitting procedure would not be under control. We thus conclude that the basic ingredients to estimate the kinematics seem to be reasonably implemented.
 
A possible explanation of the disagreement with the experiment would be 
the too small number of excited states $n_{ex}$ incorporated into the calculation. Indeed,  for $^{12}$C nucleus, $n_{ex}=5$ are included up to 16.7 MeV; for  $^{16}$O, $n_{ex}=7$ up to 19.2 MeV and $n_{ex}=7$ up to 18 MeV for $^{28}$Si. Such a reduced number of excited states may strongly affect the fragment cross-sections, more specifically the yields of light clusters with respect to the heavy ones, and the production of odd- and even-$Z$ and/or $N$ nuclei. For example there are 60 states below 8.32 MeV in $^{19}$F, 103 states below 13.97 MeV in $^{20}$Ne, 160 states below 8.19 MeV in $^{26}$Al and 62 states below 11.59 MeV in $^{28}$Si~\cite{Egidy}. Considering a small number of excited states $n_{ex}$, the code BUSCO would amplify the effect of the $Q$-values and barriers which could explain the abrupt decrease of the  cross sections of the light fragment. Addition of further excited states could be envisaged but the upper limit of the fragments to be considered in the calculation and the treatment of the fission channel are still important open questions yet to be resolved.

\subsection{Comparison with GEMINI}
In their work, N. Bohr and A.J. Wheeler~\cite{Bohr} recognized that the 
fission probability of a nucleus is governed by the number of states above the fission barrier 
and the saddle-configuration plays the role of a transition state between the CN and the scission-configuration.
Moretto~\cite{Mor2} extended this concept to the asymmetric-fission mechanism.
The GEMINI code~\cite{Charity1988} combines Hauser-Feschbach and transition-state formalisms to describe the disintegration of a hot CN 
by emission of products spanning the whole mass (charge) range from neutron to the fragment corresponding to the symmetric fission. 
The evaporation channels include $n, p, d, t$, $^{3}He$ and $\alpha$ particles. 
The emission of fragments with Z~$\ge$~3 is described within the transition-state model using the saddle conditional energy for different mass (or charge) asymmetries  deduced from the finite-range rotating liquid-drop model~\cite{Sierk}.

The decay width for the emission of a fragment ($Z,A$) from a CN at excitation energy
 $E^*_{CN}$ and spin $J$  is written as:

{
\setlength\arraycolsep{12pt}
\begin{eqnarray*}
\lefteqn{\Gamma_{Z,A}(E^*_{CN},J)} \\
 &&=\frac{1}{2\pi\rho_{0}} \int_{0}^{E^*_{CN}-E_{sad}(J)} \rho_{sad}
(U_{sad},J)d\epsilon,  
\end{eqnarray*}
}
where $U_{sad}= E^*_{CN}-E_{sad}(J)-\epsilon$ and
$\rho_{sad}$ are the thermal energy and the level density calculated at the conditional saddle-point configuration, respectively. $\epsilon$ is the kinetic energy and $E_{sad}(J)$ is the 
energy of the saddle-point configuration calculated in the finite-range liquid-drop model of Sierk. Nuclear level densities are given by the Fermi-gas formula for a fixed angular momentum $J$ as follows
\[
\rho_{sad} (U_{sad},J)\propto\frac{(2J+1)}{U_{sad}^{2}} \exp[2\sqrt{(aU_{sad})}].
\]
\begin{figure}[!b]
  \includegraphics*[scale=0.7]{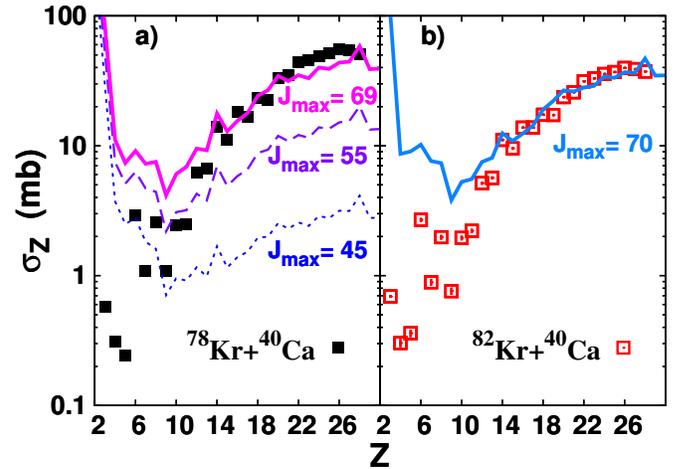}
  \caption{ (Color online) a) Experimental cross-sections for fragments emitted in the $^{78}$Kr + $^{40}$Ca reaction (full squares), compared to the predictions of the GEMINI code assuming different maximum angular momenta : $J_{max} = 69$ (thick line), $J_{max} = 55$ (dashed line) and $J_{max} = 45$ (dotted line); b) Experimental cross-sections for fragments emitted in the $^{82}$Kr~+~$^{40}$Ca reaction (open squares), compared to the predictions of the GEMINI code assuming $J_{max}$ = 70 (thick line). Calculations were performed taking $a=A/8$ MeV$^{-1}$ for the level-density parameter.}
\label{fig.CSGEMINI}
\end{figure}

In the model, the angular momentum $J_{lim}\hbar$ at which the fission barrier disappears is 69$\hbar$  for the $^{118}$Ba nucleus and 74$\hbar$  for the $^{122}$Ba nucleus. In the case of the $^{122}$Ba nucleus, $J_{lim}$ is higher than 
$J_{max}^{exp}$ deduced from data, while $J_{lim}<J_{max}^{exp}$ for the $^{118}$Ba nucleus.  
Consequently, the calculations have been performed assuming  a sharp cut-off for the angular momentum distribution with $J_{max}$~=~$J_{lim}$~=~69 for the $^{78}$Kr~+~$^{40}$Ca reaction and 
$J_{max}$~=~$J_{max}^{exp}$~=~70 for the $^{82}$Kr~+~$^{40}$Ca reaction.
Results of the calculations are reported in  Fig.~\ref{fig.CSGEMINI}a for the $^{78}$Kr~+~$^{40}$Ca system  and in
 Fig.~\ref{fig.CSGEMINI}b for the $^{82}$Kr + $^{40}$Ca reaction. As a first attempt we adopt a level-density parameter $a=A/8$ MeV$^{-1}$.
The thick line in Fig.~\ref{fig.CSGEMINI}a presents the predictions for the disintegration of $^{118}$Ba CN assuming  $J_{max}$~=~69. The shape of the $Z$-distribution for 12 $\le Z \le$ 28  is reasonably  reproduced, although the model systematically underestimates the fragment yields in the range 
18~$\le~Z~\le$~26 by roughly 20\%. A better agreement could be obtained by scaling the fission barriers but the examination of the whole $Z$-distribution is more instructive.
Indeed, the model overestimates by about a factor 10 the sum of the cross-sections for 3~$\le~Z~\le$~11. The difference comes mainly from the very high Li cross-section, while C and O calculated yields are larger by about a factor 3. To give a flavour of the $J_{max}$-dependence of the $Z$-distribution, 
results for $J_{max}= 55$ and  $J_{max}= 45$ are shown as dashed and dotted lines, respectively. C (Ne) yields are in satisfactory agreement for $J_{max}$ = 45 (55) but in both cases the whole shape is not correctly reproduced. This conclusion does not depend on the sharp cut-off approximation. Indeed, a smooth transition around $J_{max}$ would degrade the global agreement since such spin-distribution tends to depopulate the region around the symmetry and conversely to increase the yield for $Z$ around 16--20. In this way the net effect would be an increase of the width of the $Z$-distribution and thus the agreement would become worse.
Moreover, no major influence is observed by varying the level-density parameter from $A/7$ to $A/10$ MeV$^{-1}$. Regarding the staggering of the yields, one could observe a relatively good agreement above $Z$=10,  but the odd-even effect is not at all reproduced for the light fragments. The same conclusions could be written from the predictions of the disintegration of a $^{122}$Ba CN (thick line in~Fig.~\ref{fig.CSGEMINI}b). In the range 12 $\le Z \le$ 28, the model reproduces the experimental data both in shape of the $Z$-distribution and magnitude of the cross sections. As for the $^{78}$Kr~+~$^{40}$Ca reaction, the model fails to reproduce the $Z$-distribution for 3~$\le~Z~\le$~11.

The pattern of the $Z$-distributions for light fragments together with an overestimation of their yields might be due to a low barrier for mass-asymmetric fission. For  medium-mass nuclei there is  a quasi--degeneracy  of saddle- and scission-configurations, thus the total kinetic energy of the fragments is tightly related to the barrier. Considering the energy balance, a lower potential energy would correspond to higher excitation energy in the primary fragments. From the calculations, we deduced the primary $Z$-distribution before secondary decays and the multiplicity of the particles emitted from each fragments. 
A careful analysis of the results indicates that, for  3 $\le Z \le$ 11, the initial smooth behaviour of the $Z$-distribution is modified by an emission of protons and $\alpha$ particles which finally induces the fluctuations of the calculated yields shown in Fig.~\ref{fig.CSGEMINI}a,b. Thus, in the model, the fluctuations of the yields for light fragments are related to secondary emission of light particles, in contradiction with our data.  

Last, the calculated ER cross-sections $\sigma_{ER}^{GEMINI}$ for both systems (reported in Table~\ref{tab:tab1}) are in the 250-300 mb range
depending on the assumptions on level-density parameter. These values are
lower by about a factor 2 with respect to the experimental data.
The low $\sigma_{ER}^{GEMINI}$ values could be related to the mass-asymmetric barrier that
leads to enhance the light-fragment emission with respect to the \hbox{evaporation} of light particles. 

Consequently, since the $Z$-distribution mainly reflects the evolution
of the barrier profile as a function of the mass-asymmetry and angular
momentum, the  comparison with  data would indicate  a failure  of the
model  to  describe  the  boundary between  asymmetric  and  symmetric
fission  at  high angular  momentum  and  that  the landscape  of  the
potential energy surface around symmetry would be steeper than the one
implemented in  the GEMINI code.  These conclusions hold if  the decay
products are  unambiguously associated  to CN disintegration.  In this
case,  other  potential-energy  surfaces  such  as  the  one  recently
developed~\cite{LSD} might have a  better behaviour around symmetry as
indicated in a recent investigation~\cite{Mazurek}.  

\subsection{Comparison with the dinuclear system (DNS) model }

Both approaches presented in previous subsections treat the decay of an initial 
CN and disregard the collisional stage leading to its formation. 
However, a large body of data has reported on the competition between the fusion and the quasifission phenomena, the latter corresponding to 
the capture of interacting partners with a significant flow of matter and kinetic energy 
followed by a reseparation without being trapped in the CN configuration. 
For the interpretation of these two kinds of reactions, the new concept of the DNS has 
been developed and successfully compared to collisions 
involving massive nuclei~\cite{DNS}. This model has been recently applied~\cite{Kalandarov} to the decay of medium-mass excited nuclei formed 
at relatively low angular momentum. Here we compared the predictions of the DNS model 
to our data which indicate a strong relaxation at relatively high angular momentum 
and moderate excitation energy. A detailed description of the model can be found in~\cite{Kalandarov,DNS}; only the most salient features are outlined.

The DNS model describes the evolution of the interacting nuclei along two degrees of freedom; 
the relative distance $R$ between the center of the nuclei; 
the  charge and mass-asymmetry degrees of freedom, which are defined here by
the charge $Z$ and mass $A$
of the light partner of the DNS. 
After the  dissipation of kinetic energy and angular momentum 
of the relative motion, the DNS is trapped in the pocket of the interaction potential between partners. Then, a statistical equilibrium is reached in the mass-asymmetry coordinate so that the formation probability $P_{Z,A}$ of each DNS or CN configuration depends on the potential energy  $U(R_m,Z,A,J)$, calculated with respect to the potential energy of the rotational CN where $R_m$ is the location of the minimum in the interaction potential.
After the capture stage, there are nucleon drift and nucleon
diffusion  between the nuclei which constitute the DNS. Then, the excited DNS can decay with a probability $P^R_{Z,A}$ in the $R$-coordinate
if the local excitation energy of the DNS is high enough to overcome the
barrier in the nucleus-nucleus potential. Ultimately, the system evolves either towards a CN 
configuration that subsequently decays, or to a DNS configuration. The latter process, in which a 
two-body configuration is kept all along the trajectory,  is the quasifission phenomenon.

The emission probability $W_{Z,A}(E^*_{CN},J)$ of a fragment ($Z$,$A$) is calculated as the product of the DNS formation probability and the DNS decay probability:
\[
W_{Z,A}(E^*_{CN},J)=\frac{P_{Z,A} P^R_{Z,A}}{\sum_{Z',A'}P_{Z',A'}P^R_{Z',A'}},
\]
where the indexes $Z'$ and $A'$ go over all possible channels 
from the neutron evaporation to the symmetric \hbox{splitting}. 

The probability $P_{Z,A}$ is the equilibrium limit of the master equation (see~\cite{Kalandarov,DNS} for details) given by 
{
\setlength\arraycolsep{1pt}
\begin{eqnarray*}
\lefteqn{P_{Z,A}(E^*_{CN},J)}\\
&=&\frac{\exp [-U(R_m,Z,A,J)/T_{CN}(J)]}{1+\sum_{Z'=2,A'}\exp [-U(R_m,Z',A',J)/T_{CN}(J)]}.
\label{pzn1_eq}
\end{eqnarray*}}
The quasifission barrier $B^{qf}_R$, calculated as the difference between the bottom of the inner pocket and the top of the external barrier, prevents the decay of the DNS along the $R$-degree of freedom with the weight $P^R_{Z,A}$ given as 
\[
P^R_{Z,A}\sim \exp [-B_{R}^{qf}(Z,A,J)/T_{Z,A}(J)].
\]
$T_{CN}(J)$ and  $T_{Z,A}(J)$ are the  temperatures of the CN  and the
DNS,  respectively. The  Fermi-gas model  is employed  to  compute the
temperature,  with a  level-density parameter  $a$ taken  as  the high
excitation      limit     of     Ref.~\cite{level}      that     means
$a=0.114A+0.162A^{2/3}$. With this  prescription we obtained $a=17.34$
MeV$^{-1}$  for   the  $^{118}$Ba  nuclei,   equivalent  to  $a=A/6.8$
MeV$^{-1}$, a value close to those we used in BUSCO and GEMINI calculations. 

In the DNS model, all the trajectories leading to CN and QF processes represent the capture phenomenon. The pocket in the nucleus-nucleus potential disappears 
at some
critical value $J=J_{cr}$ and the DNS formation is no longer 
possible at $J>J_{cr}$. The critical value $J_{cr}$ determines the capture cross-section.
 The dominant reaction mechanism (CN or QF) strongly depends on the angular momentum. For the reactions studied here,
the driving potential at  low angular momentum shows that  CN configuration is energetically more favorable than
any DNS configuration. At higher angular momentum, the driving potential
has a minimum at the symmetric DNS and the charge (mass)-drift pushes 
the system towards symmetric configuration.  Consequently CN 
configuration becomes energetically less favorable and the
high partial waves lead to QF. However, both mechanisms coexist in a wide range of angular 
momenta. For example, 
in the case of the $^{78}$Kr~+~$^{40}$Ca reaction at  5.5 MeV/nucleon, the evaporation residue 
component accounts for about $10\%$ of the partial cross-section at $J=65$.

There are two important facets of the model. First, no $\it a$~$\it priori$ 
assumption is made on the relaxation of the $N/Z$ degree of freedom. Indeed the $N/Z$-equilibration is reached when the DNS is trapped. Secondly,
the connection between binary decay and evaporation channel 
is provided in a straightforward way by the mass-asymmetry coordinate. 
So, in the DNS model, the competition between the decay channels is treated
in a common framework. 

\begin{figure}[!b]
  \includegraphics*[scale=0.7]{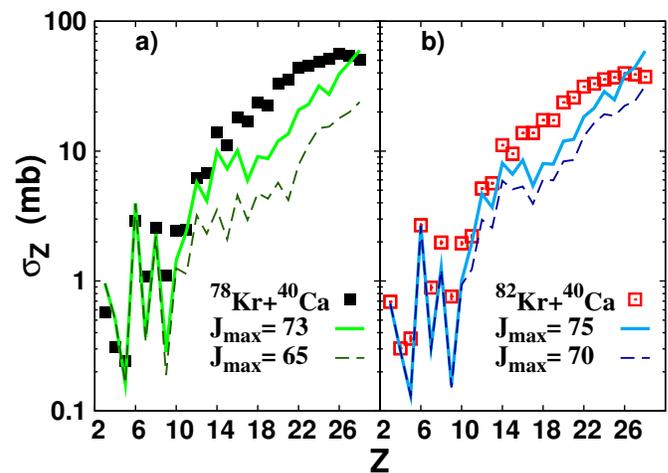} \\
 \caption{(Color online) Comparison between measured and calculated cross-sections. The calculated results with $J_{max} = 65$ ($J_{max} = 73$) for the $^{78}$Kr + $^{40}$Ca  reaction  and
 $J_{max}$ = 70 ($J_{max}$ = 75) for the $^{82}$Kr + $^{40}$Ca  reaction are shown by dashed  (solid) lines in panel a (b), respectively.
 Full (open)  squares are data from the  $^{78}$Kr + $^{40}$Ca ($^{82}$Kr + $^{40}$Ca) reaction, respectively.}
\label{fig.CSDNS}
\end{figure}

 Fig.~\ref{fig.CSDNS}a (Fig.~\ref{fig.CSDNS}b) compared DNS predictions and data for the 
 $^{78}$Kr + $^{40}$Ca ($^{82}$Kr + $^{40}$Ca) reaction, respectively.  For both reactions, the largest value of the angular momentum $J_{max}$ is taken as the critical value $J_{cr}$ according to the model. For the $^{82}$Kr + $^{40}$Ca system,  $J_{max} = 70$ is the value deduced from the measured total cross-section. Predictions with $J_{max} = 65$  for $^{78}$Kr~+~$^{40}$Ca reaction are shown for the sake of comparison. Last, the $^{8}$Be cross-section has been removed from the results of the calculations to permit the comparison with data.
   
We observe a spectacular improvement with respect to the predictions of the BUSCO and GEMINI codes. Indeed, the DNS model satisfactorily reproduces the main features of the $Z$-distributions. For both reactions, the shape of the $Z$-distributions, the strong odd-even-staggering for 5~$\le~Z~\le$~10, the small cross-sections  of light fragments as well as $\sigma_{Z}$ around $Z=28$ are well reproduced. However, for 16 $\le Z \le$ 22 the DNS model 
underestimates the fragment cross-sections by about a factor 2 to 3. Since the whole capture 
cross-section is considered, no improvement could be 
obtained within the present version of the model. Nevertheless, as reported in Table~\ref{tab:tab0}, $J_{cr}$ values of the DNS model are coherent with $l_{pocket}$ calculated using the proximity potential.  
Moreover, the ER cross-sections predicted by the DNS model $\sigma_{ER}^{DNS}$ (see Table~\ref{tab:tab1}) are compatible with data, although 
the dependence of the ER cross-section on the neutron-to-proton ratio does not follow the same trend
 as the one seen in the experiment. Thus, the depletion observed in the calculated yields for 
 16~$\le~Z~ \le$~22 might signal, in addition to the capture process, the presence of a class of deep-inelastic collisions associated to an incomplete relaxation of the entrance channel mass (charge)-asymmetry, and presumably localized in a $J$-window just above $J_{cr}$. In this case the yields of the products
near the entrance channel ($Z=20$) can exceed the predictions of the DNS model.

The staggering of the yields decreases as the atomic number increases in agreement  with the experimental findings.
Since the pairing energy of the DNS light nucleus decreases with increasing
mass number $A$, the  odd-even effect
becomes weaker  for  larger $Z$-values. Moreover, the magnitude of the staggering is also influenced by the excitation energy stored in the primary fragments (see dotted line in Fig.~\ref{fig.EsepExp}). For nuclei with $Z \lesssim 10$
the calculated average excitation energy is below the particle emission threshold and these nuclei
do not decay  further except by $\gamma$-emission which is not taken into account in the 
present version of the model.
For heavy fragments,
the average excitation energy and  spin are high enough to open-up the decay by light particles which strongly attenuates the odd-even structures of the $Z$-distributions. Such results agree with our conclusions from the analysis of the fragment-particle coincidences.

In agreement with data, $\sigma_{Z}$ for fragments
with $Z<10$ are larger for the $^{78}$Kr + $^{40}$Ca reaction. This can be explained by  their smaller mass-asymmetric decay barriers for the reaction induced with $^{78}$Kr projectile.
 
The calculated yields for 3~$\le~Z~\le$~10 show a large odd-even-staggering of about a factor 10. 
Such o-e-s is much bigger than the experimental results and is mainly due to a strong underestimation of the odd-$Z$ yields of B, N and F  while the C and O yields are well reproduced. 
 The low predicted yields of the light fragments with odd-$Z$  could be related to the prescription 
 for the static deformation for odd-nuclei which enters into the nucleus-nucleus potential.
 Reasonable changes  of static deformation would have minor effects on the yields. Another possibility would be the interplay between some 
 microscopic properties (such as pairing interaction for example) and deformation experienced by the dinuclear system en route to separation. Data would indicate an attenuation of these properties with deformation. Finally,  the nuclear level densities below separation energy could play a role in the competition between channels since they could still retain some structure behaviours which are not included in the Fermi-gas approach~\cite{Egidy}.  

\begin{figure}[!t]
\includegraphics*[scale=0.4]{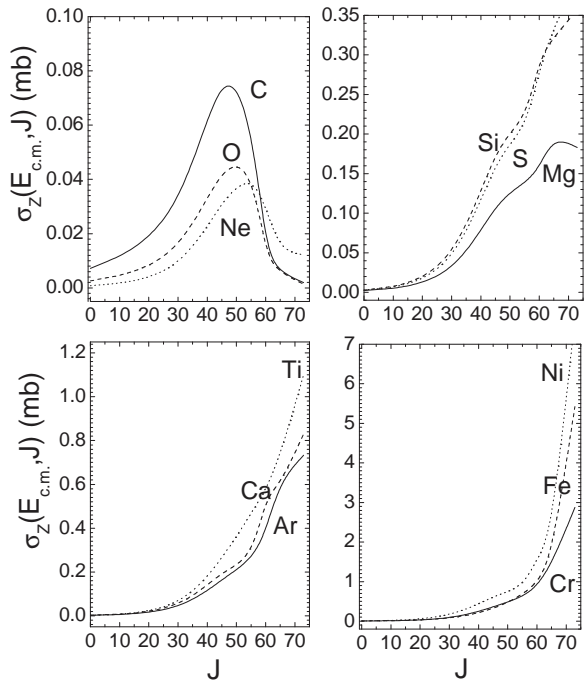} \\
\caption{ Partial cross-sections of the indicated fragments as a function of the angular momentum
for the $^{78}$Kr + $^{40}$Ca reaction at 5.5 MeV/nucleon.}
\label{fig.CSpartial}
\end{figure}

Comparing the calculated cross-sections for  $J_{max}=65$ and 73 ($J_{max}=70$ and 75)
for the $^{78}$Kr~+~$^{40}$Ca  ($^{82}$Kr~+~$^{40}$Ca) reactions (see Figs.~\ref{fig.CSDNS}a, b),
one can deduce that the contribution
from  high-partial waves to the
yields for $Z\le10$ is negligible. The calculated partial production cross-sections $\sigma_{Z}(E_{\rm c.m.},J)$ for some fragments from C to Ar are shown in Fig.~\ref{fig.CSpartial} for the $^{78}$Kr~+~$^{40}$Ca reaction at 5.5 MeV. We observed that
most of the light fragments, as for example C, O or Ne,  comes from angular momenta around
$J\hbar\approx$40--60 $\hbar$. On the contrary, 
most of the heavy fragments as for example Cr, Fe or Ni is associated to partial waves
around $J_{max}$. It is worth noting that $\sigma_{Z}(E_{\rm c.m.},J)$ develops two components for fragments with large $Z$ showing a population through both CN and quasifission mechanisms.
Examination of the results leads to the conclusion that QF is the dominant decay channel for heavy fragments while light fragments are predominantly populated by CN. Thus, the angular momentum strongly influences the
 competition between the
binary decay channels and, correspondingly, the
probability of light-fragment emission. One should also remind that the careful identification of the origin of the binary decay products  is a prerequisite before extracting information such as viscosity or fission barriers from fitting data. Thus, it would be very instructive to probe the competition between CN and QF components in the same mass region by studying small mass-asymmetric reactions where the flux going to CN is expected to dominate over a large range of incident partial waves. Experiments using a 
spin-spectrometer with high capabilities~\cite{PARIS} could be appropriate for such investigations.

The DNS model provides a good framework to describe both qualitatively and quantitatively fusion- evaporation cross-sections as well as the main features of the yields of the light fragments such as C or O.  The calculations confirm what we have deduced from the analysis of the fragment-light particle coincidences. The excitation energies and spins left in the heavy partners (Sn, Cd) after C or O  emission are very high and since these heavy nuclei are neutron-deficient, the secondary emission of light particles leads to the formation of residues  of masses A $\sim$ 100 with extremely small cross-sections. We infer that better conditions could be obtained with reactions induced by a very neutron deficient Kr beam at bombarding energy close to the Coulomb barrier~\cite{Kalandarov2}.

\section{Summary and conclusions}
We have presented the results of a study on decay modes of excited nuclei formed in $^{78,82}$Kr + $^{40}$Ca reactions at 5.5 MeV/nucleon. The $4\pi$ INDRA array which is very well suited to study the fate of violent collisions~\cite{Borderie}, has been exploited here for the first time in low bombarding energy regime. The kinetic-energy spectra, the angular distributions  and the $Z$-distribution for fragments with 3 $\le Z \le$ 28 show the characteristics of fission-like phenomenon. Analysis of the fragment-particle coincidences indicates that light partners in very asymmetric fission are produced either cold or at excitation energies  below the particle emission thresholds. We observe a persistence of structure effects from elemental cross-sections with a strong odd-even-staggering  for the lightest fragments. The magnitude of the staggering does not significantly depend on the neutron-to-proton ratio of the emitting system. The ER cross-section of the $^{78}$Kr~+~$^{40}$Ca reaction is slightly higher than the one measured in the $^{82}$Kr~+~$^{40}$Ca reaction. The fission-like component is larger by $\sim$ 25\%  for the reaction having the lowest neutron-to-proton ratio. Last, the cross sections of the light clusters (Li, Be, B) are astonishingly low.

These experimental features were compared to the predictions of various theoretical approaches assuming either the formation of CN (BUSCO, GEMINI) or describing both the collisional stage preceding the CN formation and the competition with quasifission process (DNS model). The better global agreement is obtained within the DNS framework.  
For the $^{78,82}$Kr + $^{40}$Ca reactions at 5.5 MeV/nucleon, the DNS model describes quantitatively the ER cross-sections, the odd-even-staggering of the light fragments and their low cross sections as well as a large portion of $\sigma_{Z}$ for 12 $\le Z \le$ 28. Finally, the features of the charge distribution for 3 $\le Z \le$ 28 are consistent with a strong competition between fusion-fission and quasifission processes. Examination of the results suggest that the quasifission mechanism is the dominant production mode for heavy fragments while light clusters are predominantly populated by decay of CN. 

The confrontation with data confirms the crucial role of the mass (charge)-asymmetry degree of freedom on the disintegration of excited nuclei. Moreover the potential energy surface that governs the evolution of the system must contain the contribution of microscopic properties of nuclei such as pairing interaction, shell effects or static deformations.
The interplay between the mass (charge)-asymmetry and $N/Z$-degrees of freedom and their mutual influence on the competition between fusion evaporation reactions and binary decays is yet to be explored. The advent of powerful ISOL facilities will undoubtedly provide very well adapted opportunities to bring new insights on the respective role of the mass-asymmetry and $N/Z$-degree of freedom during strongly dissipative collisions such as fusion and quasifission processes.

\section{Aknowledgments}
We thank the staff of the GANIL facility for their support during the experiment and M.~Loriggiola from LNL to have provide us with Ca targets of excellent quality. We would like to acknowledge a number of very useful discussions with R.J.~Charity, D.~Lacroix, K.~Mazurek, V.V.~Sargsyan,  and  C.~Schmitt. One of the authors (G.A) gratefully aknowledges support by a research grant from the Conseil R\'egional de Basse Normandie, France, for carrying out this work. J.P.W. is indebted to people from Dubna-JINR, INFN-Sezione di Napoli and  Universit\`a di Napoli "Federico II" for their warm hospitality. This work has been supported by  
the IN2P3-JINR,  MTA-JINR, and Polish-JINR  cooperation.

\end{document}